\documentclass{article}
\usepackage{smc}
\usepackage{times}
\usepackage{ifpdf}
\usepackage[english]{babel}
\usepackage{cite}
\usepackage[tracking=true]{microtype}
\usepackage{siunitx}

\def\papertitle{SpecSinGAN: Sound Effect Variation Synthesis Using Single-Image GANs}
\def\firstauthor{Adri{\'a}n Barahona-R{\'i}os}
\def\secondauthor{Tom Collins}
\newif\ifpdf
\ifx\pdfoutput\relax
\else
   \ifcase\pdfoutput
      \pdffalse
   \else
      \pdftrue
\fi

\ifpdf % compiling with pdflatex
  \usepackage[pdftex,
    pdftitle={\papertitle},
    pdfauthor={\firstauthor, \secondauthor},
    bookmarksnumbered, % use section numbers with bookmarks
    pdfstartview=XYZ % start with zoom=100% instead of full screen; 
   ]{hyperref}
  \usepackage[pdftex]{graphicx}
  \graphicspath{{./figures/}}
  \DeclareGraphicsExtensions{.pdf,.jpeg,.png}

  \usepackage[figure,table]{hypcap}

\else % compiling with latex
  \usepackage[dvips,
    bookmarksnumbered, % use section numbers with bookmarks
    pdfstartview=XYZ % start with zoom=100% instead of full screen
  ]{hyperref}  % hyperrefs are active in the pdf file after conversion

  \usepackage[dvips]{epsfig,graphicx}
  \graphicspath{{./figures/}}
  \DeclareGraphicsExtensions{.eps}

  \usepackage[figure,table]{hypcap}
\fi

\hypersetup{
    colorlinks,%
    citecolor=black,%
    filecolor=black,%
    linkcolor=black,%
    urlcolor=black
}

\title{\papertitle}

 \twoauthors
   {\firstauthor\text{*}\thanks{\text{*}This work was undertaken during an internship at Sony Interactive Entertainment Europe.}} {Department of Computer Science, University of York \\ 
   Sony Interactive Entertainment Europe \\
     {\tt \href{mailto:ajbr501@york.ac.uk}{ajbr501@york.ac.uk}}}
   {\secondauthor} {Department of Music, University of York \\ %
   Music Artificial Intelligence Algorithms, Inc. \\
     {\tt \href{mailto:tomthecollins@gmail.com}{tomthecollins@gmail.com}}}

\begin{document}
\capstartfalse
\maketitle
\capstarttrue
\begin{abstract}
Single-image generative adversarial networks learn from the internal distribution of a single training example to generate variations of it, removing the need of a large dataset. In this paper we introduce SpecSinGAN, an unconditional generative architecture that takes a single one-shot sound effect (e.g., a footstep; a character jump) and produces novel variations of it, as if they were different takes from the same recording session. We explore the use of multi-channel spectrograms to train the model on the various layers that comprise a single sound effect. A listening study comparing our model to real recordings and to digital signal processing procedural audio models in terms of sound plausibility and variation revealed that SpecSinGAN is more plausible and varied than the procedural audio models considered, when using multi-channel spectrograms. Sound examples can be found at the project website.\footnote{\url{www.adrianbarahonarios.com/specsingan}}\end{abstract}

\section{Introduction}\label{sec:introduction}

With the ever-growing complexity and length of video games and related media, balancing the quality and quantity of audio content has become a bigger challenge. As players may repeat actions in games, it is common that multiple sound files are used to sound design different audio variations of a single in-game interaction. Usually, this is done to prevent listener fatigue and avoid repetition \cite{zdanowicz2019game}, as well as to mimic reality (increase verisimilitude), where two sounds are hardly identical. To provide more variation to the sonic interactions, the sound design process regularly involves the use of different sound layers for a single sound effect \cite{zdanowicz2019game}. For instance, a footstep sound effect can be broken down into the tip, the heel and the shoe fabric layers. 
Sound designers generally use audio assets from pre-recorded sound libraries and/or record the sounds on demand. Some sounds, however, may be rare, difficult or time-consuming to obtain, especially if multiple variations of them in the same style are needed. Considering some modern video games may have thousands of different sonic interactions within the game environment, the amount of assets required becomes an issue in the development process. 
 
 An alternative to using pre-recorded sound samples is the use of sound synthesis to source the sound assets. In the context of game audio, this is often called generative or procedural audio \cite{farnell2010designing}. Procedural audio usually refers to the use of real-time digital signal processing (DSP) systems such as sound synthesisers, while generative (audio) can be defined as ``algorithms to produce an output that is not explicitly defined'' \cite[p.1]{yee2018procedural}. Apart from sound synthesisers, other DSP methods involve the manipulation of audio files in order to obtain a desired effect, transforming the source asset \cite{lloyd2011sound,fagerstrom2021one}. 
 Generative and procedural audio allow the dynamic creation of sound assets on demand. However, creating or using procedural or generative audio models may be challenging for sound designers. 
 With the surge in interest in deep learning, novel sound synthesis techniques are being developed which can be seen as an alternative to traditional DSP methods such as \cite{bahadoran2018fxive}, opening the possibility to synthesise sounds that may be challenging to achieve otherwise. Deep learning systems also blur the line between generative and procedural audio, with some architectures being able to synthesise audio on-demand with low latency \cite{chang2018perceptual} or even in real-time \cite{ganis2021real}. 
 
Here, we use single-image generative adversarial networks to synthesise novel variations of a single training sound effect. Generative adversarial networks (GANs) \cite{goodfellow2014generative} are a generative modeling technique where -- typically -- two neural networks (generator and discriminator) compete against each other. The generator tries to capture the data distribution by fooling the discriminator, and the discriminator tries to distinguish between real training data and the output produced by the generator. The end goal consists of having a generator capable of producing plausible novel content with a similar distribution as the training data, and this usually involves training on large datasets over extended periods of time. On the other hand, single-image GANs, such as SinGAN \cite{shaham2019singan}, are trained on overlapping patches of a single training image, modeling its underlying internal distribution and producing novel samples with similar visual content, but without the need of a large dataset. We apply the same principle to the audio domain, resulting in an alternative way to generate sound assets.

\section{Related Work}

There are multiple studies on the use of DSP systems for the synthesis of sound effects, such as in the  synthesis of footsteps \cite{nordahl2010sound} or aeroacoustic sounds \cite{selfridge2018creating}. DSP-based synthesis of sound effects has been demonstrated to be perceptually effective when suitable sound synthesis methods are used \cite{moffat2018perceptual}. Guidelines to choose a suitable sound synthesis method to synthesise a target sound have been proposed also \cite{farnell2010designing}\cite{misra2009toward}\cite{cook2002real}. There have also been DSP systems catered to generate variations of target pre-recorded impact sounds, such as in \cite{lloyd2011sound} and, after we conducted our listening study, in \cite{fagerstrom2021one}.

Audio synthesis using deep learning, often called neural audio synthesis, is a very active field of research. These techniques usually work either directly on the time domain (waveforms) or in the frequency domain (spectrograms). 
There are multiple architectures and applications that have been explored, such as autoregressive models for waveform synthesis \cite{oord2016wavenet}, GANs for unconditional \cite{donahue2018adversarial} and conditional \cite{barahona2020synthesising} waveform synthesis or for frequency-domain conditional synthesis \cite{engel2018gansynth}\cite{nistal2020drumgan} or diffusion models for conditional and unconditional waveform synthesis \cite{kong2020diffwave}. 
Other architectures, such as Differentiable DSP (DDSP) \cite{engel2019ddsp}, incorporate DSP methods like spectral modeling synthesis \cite{serra1990spectral} into the deep learning domain.
All these architectures need a training dataset of some sort.

Single-image GANs exploit the internal statistics of a single training example to generate novel variations from it. An example of these architectures is SinGAN \cite{shaham2019singan}. SinGAN is an unconditional generative model that uses a progressive growing multi-scale approach, training a fully convolutional GAN on a different resolution at each stage. The model starts producing small-sized images, which are upsampled and fed to the next stage alongside a random noise map. SinGAN uses patch-GANs \cite{isola2017image}, training on overlapping patches of the training image at the different stages. Given the GAN receptive field is fixed with respect of the image size, the model learns to capture finer details as the training progresses. This fully convolutional design also enables image generation of arbitrary size just by changing the dimensions of the input noise maps. ConSinGAN \cite{hinz2021improved} is another single-image GAN architecture. Built upon SinGAN, the authors proposed some improvements to it, such as concurrent training of the different stages or the resizing approach when building the image ``pyramid'' for the different resolutions, reducing the number of parameters and the training time. These architectures are also capable of performing other tasks such as retargeting, animation or super-resolution. There are several other single-image architectures, such as Hierarchical Patch VAE-GAN \cite{gur2020hierarchical}, which is capable of producing not only images but videos. Other single-image methods, such as Drop the GAN \cite{granot2021drop} use patch-nearest-neighbors instead. 

Catch-A-Waveform (CAW) \cite{greshler2021catch} is a recent audio time-domain architecture inspired by single-image GANs that is capable of producing novel audio samples of arbitrary length with just 20 seconds of training data. They showcase the architecture's performance on music, speech and environmental sounds (such as applause or thunderstorm), yielding promising results. CAW is also capable of performing different tasks directly on the audio domain, such as bandwidth extension, denoising or audio inpainting. In our case, we focus the modeling at the individual sound effect level with the aim of producing novel one-shots instead of streams of audio such as music excerpts or speech.

\section{Method}

\subsection{Audio Representation}

While a 2-channel frequency-domain representation consisting of a magnitude spectrogram and instantaneous frequency (IF) has been used to achieve state-of-the-art results on GAN audio synthesis of pitched musical notes \cite{engel2018gansynth} \cite{nistal2021comparing}, recently, \cite{gupta2021signal} studied the use of 1-channel phaseless log-magnitude spectrograms as an alternative for synthesising non-harmonic sounds (such as chirps or pops), achieving better perceptual results in this context. To invert the spectrogram back to audio, they reconstruct the phase using the Phase Gradient Heap Integration (PGHI) algorithm \cite{pruuvsa2017noniterative}. Another popular phase reconstruction method is the Griffin-Lim \cite{griffin1984signal} algorithm. We tested a phaseless log-magnitude spectrogram representation with both the PGHI the Griffin-Lim algorithms and, in our preliminary tests, Griffin-Lim produced better perceptual results using a 75\% frame overlap. The FFT size choice also has a significant impact on the results, being 512 the size that produced the best consistent results in our tests. We opted then to use a phaseless log-magnitude spectrograms with a FFT size of 512, 75\% overlap, a Hanning window of the same size of the FFT and reconstructing the phase with Griffin-Lim.

Similar to \cite{le2021improving}, we use multi-channel spectrograms as the input to our model. While in \cite{le2021improving} the authors use the multi-channel spectrograms to represent the pitch and intensity of musical notes, we use them to represent the different layers of the sound effect. To be precise, multi-channel spectrograms are built by stacking multiple sound layers along the channel axis, and these layers are provided by the user to the network directly (they are not extracted programmatically from layered sounds). We use the multi-channel spectrogram term (instead of multi-layer) for consistency with the literature. The use of multi-channel spectrograms instead of single-channel spectrograms in the context of sound effects synthesis presents some benefits. For instance, they allow for some parametrisation of the sound synthesis as the layers could be synced to an animation, triggered asynchronously, have different volume from each other, etc.

To build the multi-channel spectrograms first we load and normalise the sound layers in a range of $[-1,1]$. Next, we measure the longest sound effect layer in the time-domain and zero-pad the remaining layers to this length.
Then, we transform the multiple audio layers into log-magnitude spectrograms, discarding the phase information and stacking them along the tensor channel axis as if they were different channels of an image. Finally, we normalise the spectrogram (mean 0, standard deviation 1). During inference we revert this normalisation, invert the logarithm, permute the channel and batch dimensions (so the layers appear as different sounds in a batch) and transform the spectrograms back to audio. If the training sound effect has only one layer, a single-layer spectrogram is used. Unless stated otherwise, all sound layers are mono, with a sampling rate of \SI{44.1}{\kilo\hertz}.

\subsection{Architecture} \label{arch}

\begin{figure}[!h]
\centering
\includegraphics[width=7cm]{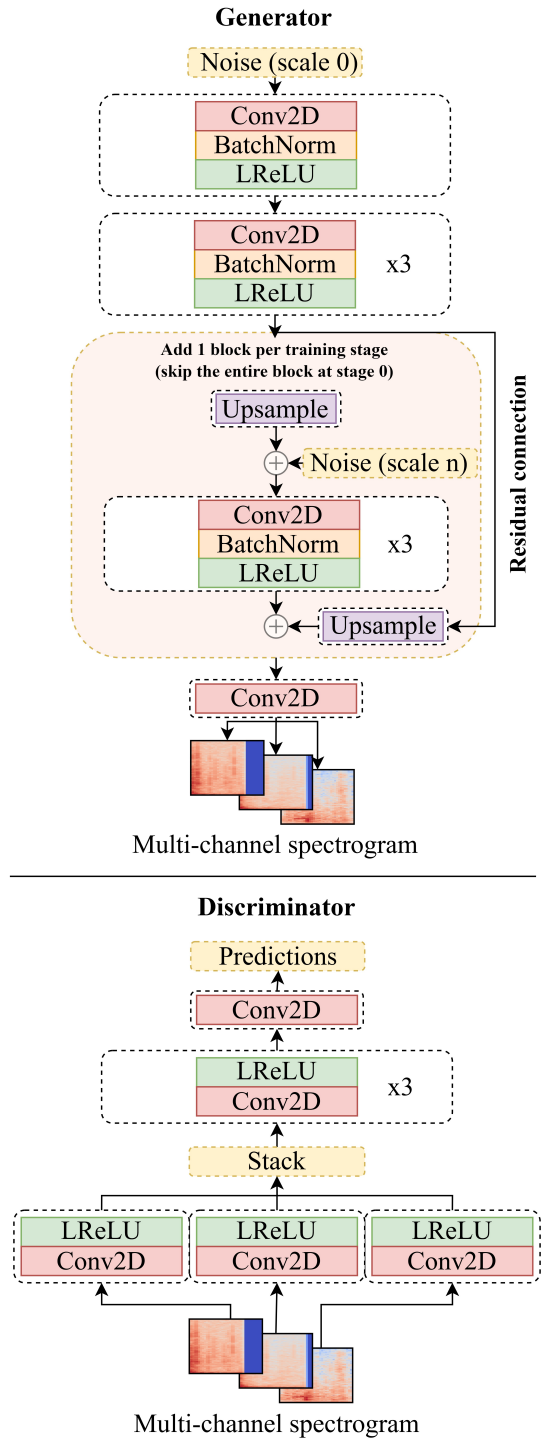}
\caption{SpecSinGAN architecture. For simplicity, the internal upsampling and padding operations in the generator are not represented in the image (details in Section \ref{arch}).}
\label{fig:arch}
\end{figure}

SpecSinGAN is built upon the ConSinGAN \cite{hinz2021improved} architecture. 
As depicted in Figure \ref{fig:arch}, the SpecSinGAN discriminator will have as many input blocks in parallel as the number of channels (sound effect layers) of the training spectrogram, each one of them consisting of a convolutional layer followed by a Leaky ReLU activacion. For instance, for a 3-channel spectrogram, the discriminator will have 3 input blocks, and each channel of the spectrogram will be processed individually by just one of them in parallel. The resulting feature maps are stacked along the batch axis, in a somewhat similar manner to \cite{le2021improving}. Then, the feature maps go onto 3 groups of convolutional layers followed by Leaky ReLU activations, and finally onto a last convolutional layer. All the convolutional layers have a kernel size of 3, stride 1 and 1 dilation rate. We use a 0.05 alpha value for the Leaky ReLU non-linearity. As suggested by the ConSinGAN authors \cite{hinz2021improved}, we implemented a second discriminator for the final training stages, finding slight improvements on the results. The second discriminator is identical to the first one, but with dilation on all its convolutional layers to increase its receptive field.

The SpecSinGAN generator is similar to the growing generator in ConSinGAN. For the first training stage, the generator has 4 convolutional layers, each one of them followed by batch normalisation and a Leaky ReLU activation, and a final output convolutional layer. As we do not constrain the range of the spectrograms, we do not use a hyperbolic tangent activation in the output of the generator. Each training stage, 3 more convolutional layers with batch normalisation and Leaky ReLU activations are added just before the output convolutional layer, increasing the generator capacity as the training progresses. All layers have the same hyperparameters as the layers in the first discriminator.

By default, we set the training stages in the training process to 10.
%10 stages in the training process.
At the start, an `image pyramid' of training spectrograms is built, going from the first stage where the images are downsampled to a coarser scale, to the last stage, where the images are at the original resolution. We set the maximum size of the image to their original resolution, and the minimum size depending on the sound.  The training starts at the coarsest scale, where the generator receives the noise maps corresponding to that stage and produces spectrograms at that size, while the discriminator classifies their different overlapping patches against the spectrogram from the training pyramid at that scale. As can be seen in Figure \ref{arch}, at the consecutive stages and until reaching the last one, the features maps from previous stages are upsampled and passed directly to the current stage, mixing them with the corresponding noise at that scale to increase diversity. We set the default noise amplification (a multiplier to weight the noise against the feature maps) to $0.1$. The feature maps from the previous stage are also upsampled and summed after the convolutional block at each stage. Overall, the training process is identical to ConSinGAN, with the exception we add a second discriminator halfway through training.

As with \cite{hinz2021improved}, we train 3 stages concurrently, using a learning rate of 0.0005 for the current stage and 0.1 scaling for the other 2 stages below. We use WGAN-GP \cite{gulrajani2017improved} for the adversarial loss, in combination with a reconstruction loss with a weight of 10. We also use the ConSinGAN upsamplig strategy for the generator, where the feature maps after the first convolutional block are slightly upsampled to increase the diversity at the edge of the images (refer to \cite{hinz2021improved} for details). More details of the training hyperparameters can be found in Section \ref{experiments}.

\begin{figure*}[t]
\centering
\includegraphics[width=14cm]{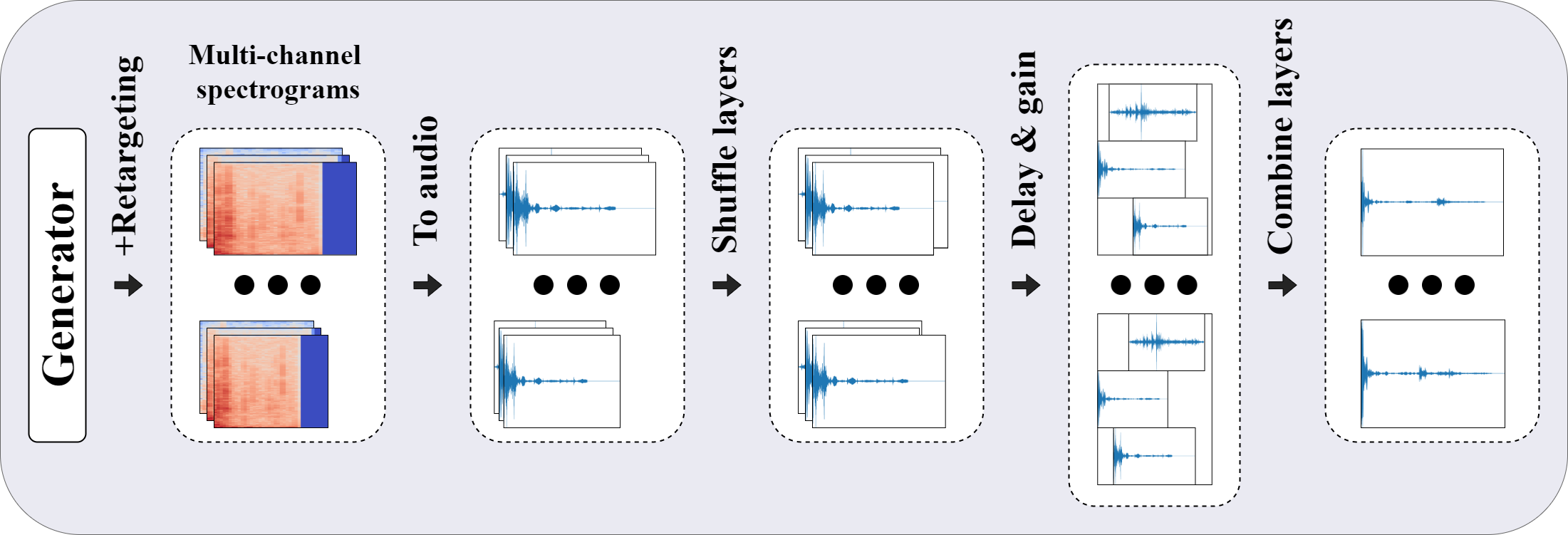}
\caption{Multi-channel synthesis during inference: the generator produces multi-channel spectrograms that are slightly different in length on their $x$-axis. They are transformed back to audio using the Griffin-Lim algorithm  \cite{griffin1984signal}, shuffling the different length layers afterwards. We finally apply a randomised delay and gain to the individual layers that comprise the sound effect, combining them to render the final audio files.}
\label{fig:multichannelsynth}
\end{figure*}

As depicted in Figure \ref{fig:multichannelsynth}, during inference we randomise the $x$-axis dimension of the noise maps that go into the generator. Note we do not randomise the $x$-axis dimension of the noise maps during training, but only during inference once the model is trained. As a result, each multi-channel spectrogram of the batch has a different length. This is thanks to the fully-convolutional nature of the network, where the input noise maps are not constrained to be of a specific shape. We transform the spectrograms into audio and shuffle the layers of the batch, combining the different-length layers. Finally, we randomise the delay and gain of the layers with respect to each other and sum the layers into a single audio file.

\section{Experiments} \label{experiments}

\begin{center}
\begin{figure*}[!h]
\centering
\includegraphics[width=17cm]{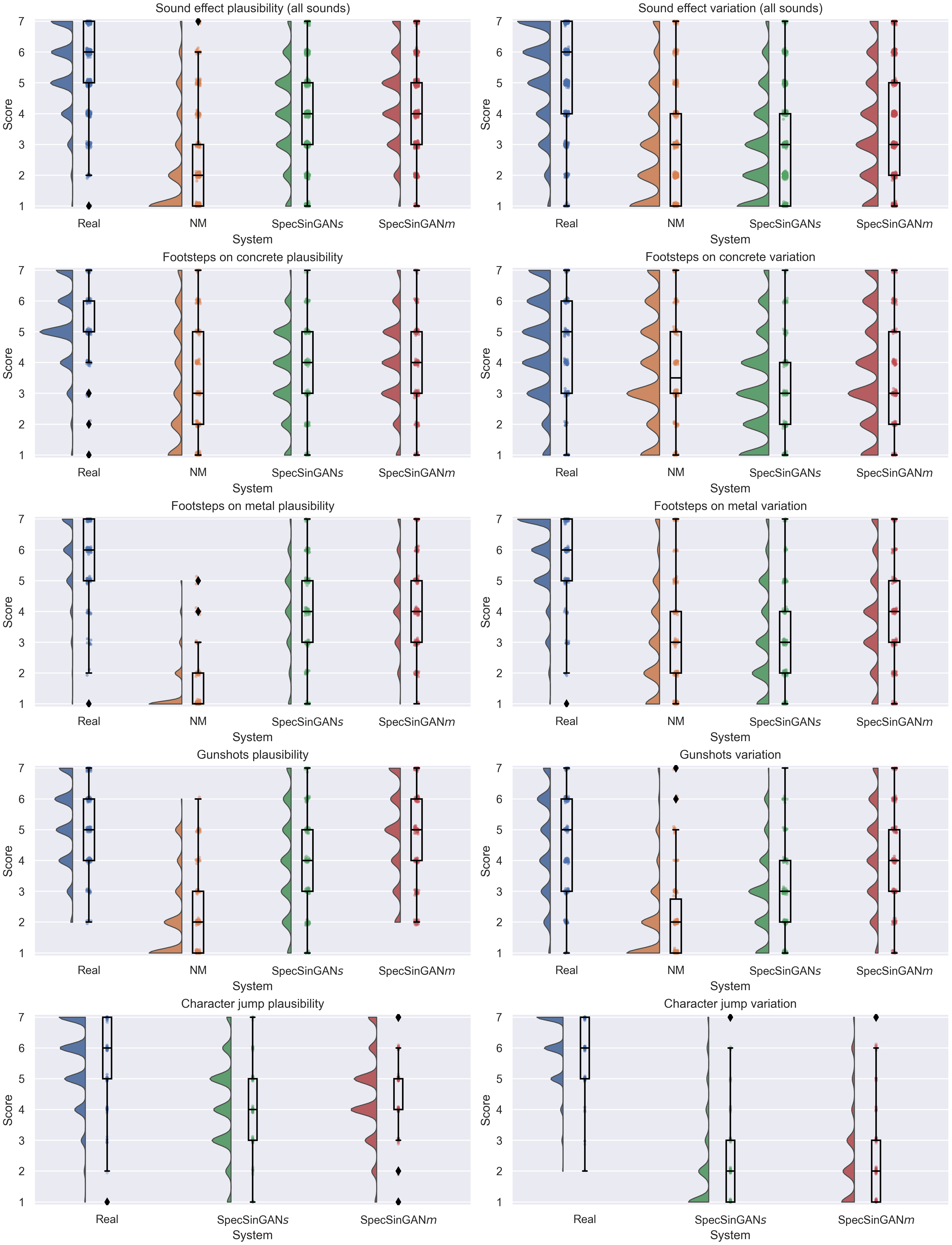}
\caption{Listening study results for the different sound effects and systems considered. Real recordings are denoted by `Real', Nemisindo by `NM' and SpecSinGAN using either single or multi-channel spectrograms by `SpecSinGAN\textsubscript{s}' and `SpecSinGAN\textsubscript{m}' respectively. Participants were asked to rate each stimuli on both plausibility and variation in a $\left[ 1..7 \right]$ scale. Note the scatter plot represents the individual ratings, with jitter added to prevent overlapping as only natural numbers were given as rating options.}
\label{fig:results_vis}
\end{figure*}
\end{center}

We selected four one-shot sound effect categories that are commonly found in video games and media, using three sound layers for each of them: footsteps on concrete (heel, tip and shoe fabric), footsteps on metal (heel, tip and metal rattle), gunshots (noise/body, mechanic component and tail) and character jump (human efforts, foley of the character clothes and metal clinks of character equipment). We collected the training sound effects from the Freesound website \cite{font2013freesound}. To assess the effect of the multi-channel spectrogram approach, we trained two models per sound effect: one with single-channel spectrograms and another with multi-channel spectrograms. The models trained with single-channel spectrograms use a single training audio file with all the different layers combined.

SpecSinGAN allows the input of both arbitrary length audio files and arbitrary numbers of layers. However, we found that different sound effects require different training hyperparameters, depending on the number of layers, the shape of the training sound, the frequency content, and the desired degree of variation. For our purposes, three layers of relatively short one-shots ($\approx$200-750 ms) are a good compromise. In general, more layers or reverberant sounds will require a higher number of iterations per training stage. For sounds with very sharp transients at the beginning, adding a small zero-padding at the start of the audio file before training can prevent artifacts. Other parameters could be changed to accommodate longer or more challenging sounds, such as the the number of training stages or even the sampling rate or the FFT size. 

In our experiments, we trained both the single and multi-channel spectrograms of the footsteps on concrete and metal for 2,000 iterations per stage, using 64 filters in the convolutional layers, a dilation rate of 3 in the second discriminator and setting the minimum tensor size of the training pyramid (on any axis) to 50. Gunshots were trained for 8,000 iterations per stage using 128 filters, a dilation rate of 2 in the second discriminator and setting the minimum size to 11. Finally, the character jumps were trained for 8,000 iterations per stage using 128 filters, a dilation rate of 3 in the second discriminator and setting the minimum size to 25. As an indication, training on an NVIDIA Tesla V100 took approximately 50 minutes for single and multi-layer footsteps on concrete and metal, 200 minutes for single-layer gunshots, 600 minutes for multi-layer gunshots and 240 and 500 minutes for single and multi-layer character jumps respectively.

During inference we randomise the delay and gain of the different layers so they are coherent with the aesthetics of the final sound effect. In terms of retargeting, we apply no more than a randomised $\pm$15\% range multiplier to the $x$-axis (corresponding with the spectrogram time axis) of the input noise maps. Multiplying the input noise maps by a larger number may result in audible artifacts. For reference, synthesising 1,000 sounds on a NVIDIA Tesla V100 took approximately 60 and 100 seconds for single and multi-channel footsteps in concrete respectively.

\section{Evaluation \& Results}

We designed a listening study to evaluate the sounds resulting from the experiments in Section \ref{experiments}, for the 4 catgeories of footsteps on concrete, metal, gunshot, and character jump. We compared 4 different systems on plausibility and variation: real recordings (hereafter, Real), SpecSinGAN with single and multi-channel spectrograms (denoted by the subscript ``s'' or “m” respectively), and Nemisindo \cite[hereafter, NM]{bahadoran2018fxive}. NM is a web-based procedural audio service that enables the creation of synthesised sound effects using DSP methods.
For the character jump sounds, we only compared Real and SpecSinGAN variants, as NM does not offer this type of sound at the time the study was carried out. 
Sounds for the Real category were sourced from the same Freesound \cite{font2013freesound} training examples used in Section \ref{experiments}, taking sound variations from the same file or designing them (e.g., cutting, equalising, fading) when needed. The SpecSinGAN sounds are those of Section \ref{experiments}. NM sounds were taken directly from their presets (without searching the parameter space), finding the closest ones to the target category of sounds.

The sounds were presented concatenated to make sound actions (e.g., footsteps turned into walking), resulting in audio clips approximately 5-seconds long. 
Participants rated 5 sounds 
per category per system in terms of plausibility and variation on a scale of 1 (not at all plausible/varied) to 7 (completely plausible/appropriately varied).
We used the Prolific platform to conduct the listening test and recruited 30 participants, compensating them £9/h. We pre-screened participants such that only those over 18 years old who play video games for at least 6 hours a week were selected, and encouraged them to use headphones during the evaluation. The listening study results are shown in Figure \ref{fig:results_vis}.

We used non-parametric Bayes factor analysis (BFA) to investigate our hypotheses about how the systems would compare in a listening study. This is because the rating data cannot be assumed to be normally distributed, and because Bayesian hypothesis testing is widely regarded as superior to the frequentist variety, with the former allowing for finding evidence in favour of ``no difference between systems'' if the data suggest as much \cite{van2020bayesian}. We hypothesised that Real would have better plausibility and variation than any other system, and the BFA found extreme evidence for this ($BF_{10} > 100$). We also hypothesised SpecSinGAN would have slightly more plausibility than NM for footsteps on concrete, finding anecdotal and strong evidence for this for SpecSinGAN\textsubscript{s} ($BF_{10} = 1.09$) and SpecSinGAN\textsubscript{m} ($BF_{10} = 12.5$) respectively. In addition, we hypothesised that SpecSinGAN would have higher plausibility ratings than NM for footsteps on metal, and the BFA found extreme evidence for this ($BF_{10} > 100$). 
Finally, we hypothesised SpecSinGAN and NM would have similar plausibility for gunshots, and the BFA rejected this ($BF_{10} > 100$). The data suggested SpeSinGAN having more plausibility than NM for gunshots. Regarding variation, we hypothesised SpecSinGAN and NM would have similar values, confirmed for SpecSinGAN\textsubscript{s} according to the BFA ($BF_{10} = 0.09$) and rejected for SpecSinGAN\textsubscript{m} ($BF_{10} > 100$). In the latter case, the data suggest SpecSinGAN\textsubscript{m} had higher variation values than NM.

\section{Discussion}
Audio asset creation can be a time-consuming process. In this paper we presented SpecSinGAN, an unconditional generative architecture capable of synthesising novel variations of one-shot sound effects with a single training example. 

SpecSinGAN performed statistically significantly better in the listening study, compared to the procedural audio models considered.
We used the NM presets available at the time of the listening study instead of searching the parameter space, so, while this is a reasonable choice, it is possible that NM (or any other DSP tool) would produce better results following a more exhaustive search of its parameter space. 

In future work, we plan to introduce user control over the synthesis, alongside increasing its plausibility and variation to be on par with real recordings. Another improvement would be the implementation of automatic hyperparameter tuning, given that, as discussed in Section \ref{experiments}, different sounds required different hyperparameters and, despite there being some intuitions on how to tune them, this still involves a manual process. 

We would also observe that, despite showing SpecSinGAN is a viable alternative to synthesise arbitrary one-shot sound effects, DSP-based systems are also capable of producing continuous streams of audio, as well as running in real-time with direct input from either human-interpretable controls or in-game parameters, granting them great adaptability. We also acknowledge that, while we focused on arbitrary sound effects, further listening studies need to be carried out to understand how SpecSinGAN compares to DSP methods such as \cite{fagerstrom2021one} for generating variations of target percussive sounds and to \cite{greshler2021catch}, adapting it to work with shorter sounds at \SI{44.1}{\kilo\hertz}. 
We suggest, however, that SpecSinGAN can be useful in a contexts where 1)~sound designers need to produce novel variations of a specific pre-recorded sound, or 2)~data is scarce, in which case SpecSinGAN acts as a data augmentation tool.

\begin{acknowledgments}
This work was supported by the EPSRC Centre for Doctoral Training in Intelligent Games \& Game Intelligence (IGGI) [EP/L015846/1] and Sony Interactive Entertainment Europe.\end{acknowledgments} 

%\narrowstyle

%%%%%%%%%%%%%%%%%%%%%%%%%%%%%%%%%%%%%%%%%%%%%%%%%%%%%%%%%%%%%%%%%%%%%%%%%%%%%
%bibliography here
\bibliography{smc2022template.bib}

\end{document}